\documentstyle[preprint,aps]{revtex}
\tightenlines
\def\Journal#1#2#3#4{{#1} {\bf #2}, #3 (#4)}

\def\CMP{{\em Commun. Math. Phys.}}
\def\IJMPA{{\em Int. J. Mod. Phys.} A}
\def\IJMPB{{\em Int. J. Mod. Phys.} B}
\def\MPLA{{\em Mod. Phys. Lett.} A}

\def\NPB{{\em Nucl. Phys.} B}

\def\PLB{{\em Phys. Lett.} B}

\def\PRA{{\em Phys. Rev.} A}

\def\PRD{{\em Phys. Rev.} D}

\def\PREPC{{\em Phys. Rep.} C}
\def\RMP{{\em Rev. Mod. Phys.}}

\def\ZPC{{\em Z. Phys.} C}

\newcommand{\be}{\begin{equation}}
\newcommand{\ee}{\end{equation}}
\newcommand{\bea}{\begin{eqnarray}}
\newcommand{\eea}{\end{eqnarray}}

\newcommand{\hf} {{1\over2}}
\newcommand{\nonu}{\nonumber\\}
\def\dk{\Delta k}

\def\la{\lambda}
\def\eq#1{(\ref{#1})}
\def\cm{{\cal M}}
\def\ord{{\cal O}}

\begin{document}
\title{Functional Callan-Symanzik equation}

\author{Jean Alexandre\thanks{alexandr@lpt1.u-strasbg.fr}}
\address{Laboratory of Theoretical Physics, Louis Pasteur University\\
3 rue de l'Universit\'e 67087 Strasbourg, Cedex, France}

\author{Janos Polonyi\thanks{polonyi@fresnel.u-strasbg.fr}}
\address{Laboratory of Theoretical Physics, Louis Pasteur University\\
3 rue de l'Universit\'e 67087 Strasbourg, Cedex, France\\
and\\
Department of Atomic Physics, L. E\"otv\"os University\\
P\'azm\'any P. S\'et\'any 1/A 1117 Budapest, Hungary}
\date{\today}
\maketitle
\begin{abstract}
We describe a functional method to obtain the exact evolution equation of the
effective action with a parameter of the bare theory. When this parameter
happens to be the bare mass of the scalar field, we find a functional 
generalization of the Callan-Symanzik equations. Another possibility
is when this parameter is the Planck constant and controls the
amplitude of the fluctuations. We
show the similarity of these equations with the Wilsonian renormalization
group flows and also recover the usual one loop effective action.
\end{abstract}

\vfil
\eject

\section{Introduction}

The renormalization group originally provides 
an insight into the scale dependence
of the coupling constants \cite{wilsrg}. Another, more recent use
is to perform a partial resummation of the perturbation 
expansion by making an infinitesimal change of the cutoff 
in a time and using the functional formalism \cite{wh}-\cite{tetra}.
Both goals are realized by the blocking procedure, the 
successive elimination of the degrees of freedom 
which lie above the running ultraviolet cutoff. The resulting
evolution equation yields the dependence of the coupling 
constants on the cutoff. This procedure suffers from consistency
problems with the gradient expansion if we consider a sharp cutoff.
To avoid it one usually introduces a smooth cutoff, but its form
is not unique and the physical implications are not clear either.
Moreover,
in the framework of gauge theories, the slicing procedure in momentum space 
is not gauge invariant. 

The "exact" renormalizaton program \cite{polc}-\cite{tetra}, with the setting
up of the flow equations for the coupling constants, helps to compute the
effective action of the theory in the limit where the cutoff goes to zero.
But these flow equations suffer from the same problems and it is not
clear if the resulting effective action is free of these inconsistencies.

We propose here another way of computing the effective action of a given
bare theory, by introducing a control parameter $\lambda$
in the theory and looking at
the evolution of the effective action (defined, as usual, by the Legendre
transform of the generator functional for the connected graphs)
in this parameter. The parameter in question is chosen in such a manner that
for its large values the fluctuations are suppressed and for small values the
original bare theory is recovered. It is obvious from such a general
setting that the usual "exact" evolution equations can be recovered
and the former represents a certain generalization of the latter.

Interesting generalizations are the choices (i) $\lambda=M^2$, of the mass,
or (ii) $\lambda=\hbar^{-1}$, the inverse of the Planck constant
as the control parameter. The case (i) yields the
functional generalization of the Callan-Symanzik equation. The
choice (ii) produces a manifestly gauge invariant evolution equation
for gauge models and represents a "renormalization group improved"
loop-expansion scheme. Note that in both cases $\lambda$ controls
the amplitude of the fluctuations and the "evolution" in $\lambda$
is the resummation of the effects of fluctuations with growing
amplitude.

An essential difference between these schemes and the "exact" renormalization 
group
procedure is that the control parameter of the latter performs the
role of the cutoff. In fact, the influence of $\lambda$ on the dynamics
is momentum independent in the case (i) and has a weak momentum
dependence in (ii) thereby a UV cutoff is required in the model.
Since the evolution equation is obtained by means of the regulated,
well defined path integral the scheme is genuinely non-perturbative,
as for lattice regularization. Naturally the question of the convergence when 
one
attempts to remove the cutoff after having integrated the evolution equation 
is rather involved and we cen offer no new results compared to the other
regularization schemes. 

Different standard methods based on the multiplicative renormalization scheme
are briefly discussed in Section II. Our evolution equation
of the effective action in control parameter which corresponds to
a quadratic term in the bare action is introduced in Section III.
Section IV contains the description of case (i) above, the
application of our strategy to obtain the functional generalization
of the Callan-Symanzik equation. The possibility of resumming the
loop expansion by solving the evolution equation, case (ii) mentioned above,
is shown in Section V. We present the procedure for a scalar model only but
it is obvious from the construction that, though technically more
involved due to the index structure, the generalization of this scheme
is trivial for gauge models. The Section VI is for the summary.
The appendices
give details about the Legendre transformation and the gradient expansion.

\section{Renormalization group schemes}
The traditional field theoretical methods for
the renormalization group equation are based on the 
simplification offered by placing the
ultraviolet cutoff far away from the scale of the observables.
Such a separation of the scales removes the non-universal pieces of the
renormalized action and the rather complicated blocking 
step can be simplified by retaining the renormalizable coupling constants
only. The underlying formalism is the renormalized perturbation expansion,
in particular the multiplicative renormalization scheme.
The usual perturbative proof of the renormalizability
asserts that the renormalized field and the Green
functions can be written in terms of the bare quantities as
\be
G_n(p_1,\cdots,p_n;g_R(\mu))_R=Z^{-{n\over2}}
\left(g_R,g_B,{\Lambda\over\mu}\right)
G_n(p_1,\cdots,p_n;g_B,\Lambda)_B
\left(1+O\left({p^2\over\Lambda^2}\right)\right),
\label{mult}
\ee
where $\Lambda$ is the cutoff and the renormalized coupling 
constants are defined by some renormalization conditions imposed
at $p^2=\mu^2$. The evolution equation for the
bare and the renormalized coupling constants result from the
requirements
\be\label{barerg}
{d\over d\Lambda}G_R={d\over d\Lambda}Z^{-{n\over2}}G_B=0,
\ee
\be\label{renrg}
{d\over d\mu}G_B={d\over d\mu}Z^{n\over2}G_R=0.
\ee
Note that the non-renormalizable operators can not be treated
in this fashion because the $O(p^2/\Lambda^2)$ contributions
are neglected in \eq{mult}. The renormalization of 
composite operators and the corresponding operator
mixing requires the introduction of additional terms 
in the lagrangian. Another aspect of this shortcoming is 
that these methods are useful for the study of the ultraviolet scaling
laws only. The study of the infrared scaling or models where there 
are several non-trivial scaling regimes \cite{glob} require the more 
powerful functional form, introduced below.

Another conventional procedure is the Callan-Symanzik 
equation which is based on the
change of the bare mass parameter,
\be
{d\over dm^2}G_B={d\over dm^2}Z^{n\over2}G_R=Z^{n\over2}Z_{\phi^2}G^{comp}_R
\ee
where $Z_{\phi^2}$ is the renormalization constant for the composite operator
$\phi^2(x)$ and $G^{comp}$ is the Green function with an additional
insertion of $\phi^2(p=0)$. One can convert the mass
dependence inferred from the Callan-Symanzik equation
into the momentum dependence by means of dimensional analysis 
and the resulting expression is usually called a
renormalization group equation. 

The functional generalizations of the renormalization group method
which are based on the infinitesimally small change of the cutoff
allows up to follow the mixing of non-renormalizable operators,
as well, and to trace the evolution close to the cutoff. Another
advantage of these methods is that the renormalization group
equation is either exact or holds in every order of the
loop expansion. In the framework of the blocking transformations \cite{wilsrg},
the original work by Wegner and Houghton \cite{wh} describes the evolution of 
the effective action with the scale $k$ obtained by eliminating the degrees 
of freedom with momenta $|p|>k$. Starting from a bare theory with cutoff
$k=\Lambda$, the successive 
lowering of the cutoff $k\to k-\Delta k$ gives the evolution equation
in dimension $d$
\be\label{wheq}
S_{k-\delta k}[\phi]=S_k[\phi]+\frac{1}{2}\mbox{tr}\ln\left(\frac{\delta^2 S_k}
{\delta\phi(p)\delta\phi(p)}\right)+{\cal O}\left(\frac{\Delta k}{k}\right)^2
\ee
where the trace is to be take over momenta $p\in[k-\Delta k,k]$ and
therefore is of order $\Delta k/k$. If we make the local potential approximation
as a functional ansatz for the running action, i.e. we take for any $k$
\be
S_k[\phi]=\int 
dx\left[\frac{1}{2}\partial_\mu\phi_x\partial_\mu\phi_x+U_k(\phi)\right]
\ee
we then obtain for the O(N) model in the limit $\Delta k/k\to 0$ the partial 
differential 
equation 
\be\label{whlocpot}
k\partial_kU_k(\rho)=-\frac{k^d\Omega_d}{2(2\pi)^d}\ln\left[
\left(\frac{k^2+\partial_\rho^2 U_k(\rho)}{k^2+\partial_\rho^2 
U_k(\rho_0)}\right)
\left(\frac{k^2+\frac{1}{\rho}\partial_\rho U_k(\rho)}
{k^2+\frac{1}{\rho_0}\partial_\rho U_k(\rho_0)}\right)^{N-1}\right] 
\ee
where $\rho$ is the modulus of the $N$-component field and $\Omega_d$ the solid 
angle 
in dimension $d$.

We note the existence of 
other works based on a smooth cutoff procedure \cite{polc}-\cite{tetra} but  
introducing the mass scale in a similar manner.

The method we will present in this paper proposes another approach of the 
evolution of the effective action with a mass scale. The idea will be to 
develop a
functional extension of the Callan-Symanzik procedure. Starting from our 
equations,
we will make the analogy with the exact Wilsonian renormalization group and also
we will find the usual well-known one loop effective action.
The functional equations we obtain are exact and can be used 
as an alternative method to compute
the effective action.

\section{Evolution equation}

Our goal is to obtain the effective action $\Gamma[\phi]$
of the Euclidean model defined by the action $S_B[\phi]$.
The usual Legendre transformation yields
\be
e^{W[j]}=\int D[\phi]e^{-S_B[\phi]+\int_xj_x\phi_x}
\ee
and
\be
W[j]+\Gamma[\phi]=\int_xj_x\phi_x=j\cdot\phi,
\ee
the source $j$ is supposed to be expressed in terms of 
\be
\phi_x={\delta W[j]\over\delta j_x}.
\label{phij}
\ee
A cutoff $\Lambda$ is assumed implicitly in the path integral and 
$S_B[\phi]$ stands for the bare, cutoff action. 

We modify the bare action
\be
S_B[\phi]\longrightarrow S_\lambda[\phi]=\lambda S_s[\phi]+S_B[\phi]
\ee
in such a manner that the model with $\la\to\infty$ be soluble,
because the path integral for $\la=\infty$ 
contains no fluctuations. The role of the new piece in the action
is to suppress the fluctuations around the minimum $\phi_\infty$ of the action
for large value of $\lambda$ and render the model perturbative. 

We plan to follow the $\la$ dependence of the effective action
by integrating out the functional differential equation
\be
\partial_\la\Gamma={\cal F}_\la[\Gamma]\label{evol}
\ee
from the initial condition 
\be
\Gamma_{\la_{init}}[\phi]=\la_{init}S_s[\phi]+S_B[\phi],
\ee
imposed at $\la_{init}\approx\infty$ to $\la=0$. \eq{evol} can be
interpreted as a generalization of the
Callan-Symanzik equation because both generate
a one-parameter family of different theories organized according
to the strength of the quantum fluctuations
\footnote{Note that the inverse mass is proportional to the 
amplitude of the fluctuations.}. So long 
as the parameter $\la$ introduces a renormalization scale, $\mu(\la)$,
the trajectory $\Gamma_{\la(\mu)}[\phi]$ in the effective action
space can be thought as a renormalized trajectory. Another
way to interpret \eq{evol} is to consider its integration as a
method which builds up the fluctuations of the model with 
$\la=\infty$ by summing up the effects of increasing the fluctuation 
strength infinitesimally, $\la\to\la-\Delta\la$. 
The gradient expansion
is compatible with \eq{evol} if the suppression is sufficiently
smooth in the momentum space, i.e. $S_s$ is a local functional.

The starting point to find ${\cal F}_\la[\Gamma]$ is
the relation
\be
\partial_\la\Gamma[\phi]=-\partial_\la W[j]
-\frac{\delta W[j]}{\delta j}.\partial_\lambda j
+\partial_\lambda j.\phi=-\partial_\la W[j],
\ee
$\lambda$ and $\phi$ being the independent variables. This relation
will be used together with
\be
\partial_\la W[j]=-e^{-W[j]}\int D[\phi]S_s[\phi]
e^{-\la S_s[\phi]-S_B[\phi]}=-e^{-W[j]}
S_s\left[{\delta\over\delta j}\right]e^{W[j]}.
\ee
It is useful to perform the replacement
\be
\Gamma[\phi]\longrightarrow\la S_s[\phi]+\Gamma[\phi]
\ee
which results the evolution equation
\be
\partial_\la\Gamma[\phi]=e^{-W[j]}
S_s\left[{\delta\over\delta j}\right]e^{W[j]}-S_s[\phi].
\label{evolk}
\ee

The next question is the choice of the 
fluctuation-suppressing action $S_s[\phi]$. 
The simplest is to use a quadratic suppression term,
\be
S_s[\phi]=\hf\int_{x,y}\phi_x\cm_{x,y}\phi_y
=\hf\phi\cdot\cm\cdot\phi.
\label{qsup}
\ee
We discuss in section V the choice $S_s=S_B$ and look at the evolution of 
the effective action with $\hbar$. 
The evolution equation \eq{evolk} there sums up the loop expansion and
produces the dependence in $\hbar$. This particular choice will be motivated 
by the extension of the present work to gauge theories.

We return now to the case of a simple
scalar field without local symmetry, \eq{qsup}. The 
corresponding evolution equation can be ontained from \eq{evolk},
and considering the relation \eq{propi} between the functional derivatives of
$W[j]$ and $\Gamma[\phi]$,
\bea\label{evolkk}
\partial_\la\Gamma[\phi]&=&\hf\int_{x,y}\cm_{x,y}\left[
W^{(2)}_{x,y}+\phi_x\phi_y\right]
-\hf\int_{x,y}\phi_x\cm_{x,y}\phi_y\nonu
&=&\hf\int_{x,y}\cm_{x,y}\left[\Gamma^{(2)}_{x,y}+\la\cm_{x,y}
\right]^{-1}
\eea
where the functional derivatives are denoted by
\be
\Gamma^{(n)}_{x_1,\cdots,x_n}=
{\delta^n\Gamma[\phi]\over\delta\phi_{x_1}\cdots\delta\phi_{x_n}}.                     
\ee                                                                                    
\eq{evolkk} reads in an operator notation
\be
\partial_\la\Gamma[\phi]=\hf\mbox{Tr}\left\{\cm\cdot\left[
\la\cm+\Gamma^{(2)}\right]^{-1}\right\}, 
\ee
We should bear in mind that $\Gamma^{(n)}_{x_1,\cdots,x_n}$ 
remains a functional of the field $\phi_x$. 

It is illuminating to compare
this result with the evolution equations presented in refs.
\cite{polc}-\cite{tetra} in the framework of Wilsonian renormalization
group equations
\be
\partial_k\Gamma[\phi]=\hf\mbox{Tr}\left\{\partial_kG^{-1}_k\cdot\left[
G^{-1}_k+\Gamma^{(2)}\right]^{-1}\right\}.
\label{pwm}
\ee
where the role of $S_s^{(2)}$ is played here by the propagator $G_k(p)$ which 
contains the scale parameter $k$ such that the fluctuations with momenta $|p|>k$
are suppressed.
The formal similarity with \eq{evolkk}
reflects that the different schemes agree
in "turning on" the fluctuations in infinitesimal steps. We will come back to 
this 
remark in section IV.

The evolution equation can be converted into
a more treatable form by the means of the gradient
expansion,
\be\label{gradexp}
\Gamma[\phi]=\int_x\left\{\frac{1}{2}Z_x
(\partial_\mu\phi_x)^2+U_x+O(\partial^4)\right\}
\ee
where the notation $f_x=f(\phi_x)$ was introduced. This ansatz gives
\bea
\Gamma^{(1)}_{x_1}&=&-\hf Z^{(1)}{x_1}(\partial_\mu\phi_{x_1})^2
-Z_{x_1}\Box\phi_{x_1}+U^{(1)}_{x_1}\\
\Gamma^{(2)}_{x_1,x_2}
&=&-\hf\delta_{x_1,x_2}Z^{(2)}_{x_1}(\partial_\mu\phi_{x_1})^2
-\partial_\mu\delta_{x_1,x_2}Z^{(1)}_{x_1}\partial_\mu\phi_{x_1}\nonu
&&-\delta_{x_1,x_2}Z^{(1)}_{x_1}\Box\phi_{x_1}
-\Box\delta_{x_1,x_2}Z_{x_1}+U^{(2)}_{x_1}\nonumber
\eea
where the $f^{(n)}(\phi)=\partial^n_\phi f(\phi)$.
Such an expansion is unsuitable for $W[j]$ due to the strong
non-locality of the propagator but might be more
successful for the effective action where the one-particle
irreducible structure and the removal of the propagator at the 
external legs of the contributing diagrams strongly reduce
the non-local effects. The replacement of this ansatz into 
\eq{evolk} gives (c.f. Appendix B.)
\bea\label{evoluz}
\partial_\la U_\la(\phi)&=&\hf\int_p
{{\cal M}(p)\over\la {\cal M}(p)+Z_\la(\phi)p^2+U_\la^{(2)}(\phi)}\nonu
\partial_\la Z_\la(\phi)&=&\hf\int_p
{\cal M}(p)\left[-\frac{Z_\la^{(2)}(\phi)}
{\left(\la {\cal M}(p)+Z_\la(\phi)p^2+U_\la^{(2)}(\phi)\right)^2}\right.\nonu
&+&2Z_\la^{(1)}(\phi)\frac{2\left(Z_\la^{(1)}(\phi)p^2+U_\la^{(3)}(\phi)\right)
+Z_\la^{(1)}(\phi)p^2/d}
{\left(\la {\cal M}(p)+Z_\la(\phi)p^2+U_\la^{(2)}(\phi)\right)^3}\nonu
&-&\frac{\left(Z_\la^{(1)}(\phi)p^2+U_\la^{(3)}(\phi)\right)^2
\left(\la\Box {\cal M}(p)+2Z_\la(\phi)\right)}
{\left(\la {\cal M}(p)+Z_\la(\phi)p^2+U_\la^{(2)}(\phi)\right)^4}\nonu
&-&\frac{4}{d}Z_\la^{(1)}(\phi)
\left(Z_\la^{(1)}(\phi)p^2+U_\la^{(3)}(\phi)\right)
\frac{\left(\la p_\mu \partial_\mu {\cal M}(p)+2Z_\la(\phi)p^2\right)}
{\left(\la {\cal M}(p)+Z_\la(\phi)p^2+U_\la^{(2)}(\phi)\right)^4}\nonu
&+&\left.\frac{2}{d}\frac{\left(Z_\la^{(1)}(\phi)p^2+U_\la^{(3)}(\phi)\right)^2
\left(\la\partial_\mu {\cal M}(p)+2Z_\la(\phi)p_\mu\right)^2}
{\left(\la {\cal M}(p)+Z_\la(\phi)p^2+U_\la^{(2)}(\phi)\right)^5}\right]
\eea
where $\int_p=\int{d^dp\over(2\pi)^d}$ and we assumed that $\partial_\mu\cm(p)$
is proportional to $p_\mu$.

\section{Mass dependence}
We take $\la=m^2$ with
\be
\cm_{x,y}=\delta_{x,y}
\ee
which minimizes strength of the higher order derivative terms generated
during the evolution by being a momentum independent suppression mechanism.
The evolution equation
is the functional differential renormalization group version 
of the Callan-Symanzik equation,
\be
\partial_{m^2}\Gamma[\phi]=\hf\mbox{Tr}\left[m^2\delta_{x,y}
+\Gamma^{(2)}_{x,y}\right]^{-1}.
\ee
The projection of this functional equation onto the gradient expansion 
ansatz gives
\bea\label{csrg}
\partial_{m^2}U(\phi)&=&\hf\int_p
{1\over Z(\phi)p^2+m^2+U^{(2)}(\phi)}\nonu
\partial_{m^2}Z(\phi)&=&\hf\int_p
\left[-\frac{Z^{(2)}(\phi)}
{\left(Z(\phi)p^2+m^2+U^{(2)}(\phi)\right)^2}\right.\nonu
&~&~~~~~~~+2Z^{(1)}(\phi)\frac{p^2/dZ^{(1)}(\phi)+
2\left(Z^{(1)}(\phi)p^2+U^{(3)}(\phi)\right)}
{\left(Z(\phi)p^2+m^2+U^{(2)}(\phi)\right)^3}\nonu
&~&~~~~~~~-2Z(\phi)\frac{\left(Z^{(1)}(\phi)p^2+U^{(3)}(\phi)\right)^2}
{\left(Z(\phi)p^2+m^2+U^{(2)}(\phi)\right)^4}\nonu
&~&~~~~~~~-\frac{8p^2}{d}Z(\phi)Z^{(1)}(\phi)
\frac{\left(Z^{(1)}(\phi)p^2+U^{(3)}(\phi)\right)}
{\left(Z(\phi)p^2+m^2+U^{(2)}(\phi)\right)^4}\nonu
&~&~~~~~~~+\left.\frac{8p^2}{d}Z^2(\phi)\frac{\left(Z^{(1)}(\phi)p^2+U^{(3)}
(\phi)\right)^2}
{\left(Z(\phi)p^2+m^2+U^{(2)}(\phi)\right)^5}\right]
\eea
It is important to bear in mind that we are dealing here
with a well regulated theory and that the procedure described here 
does not aim at removing the cutoff $\Lambda$ which remains an
essential parameter. 

Let us now simplify the differential equation for $U(\phi)$ and 
$Z(\phi)$ by integrating over $p$ in \eq{csrg} with sharp
momentum cutoff $\Lambda$\footnote{
The sharp momentum space cutoff generates nonlocal interactions.
Since these nonlocal contributions come from the surface terms of the loop
integrals they are suppressed in a renormalizable theory when $\Lambda$
is kept large. Thus the gradient expansion ansatz can be justified
for the evolution \eq{csrg}.}
in four dimensions,
\bea\label{diffuz}
\partial_{m^2}U(\phi)&=&\frac{1}{32\pi^2Z(\phi)}
\left[\Lambda^2-{m^2+U^{(2)}(\phi)\over Z(\phi)}
\ln\left(1+\frac{Z(\phi)\Lambda^2}{m^2+U^{(2)}
(\phi)}\right)\right]\nonu
\partial_{m^2}Z(\phi)&=&\frac{1}{32\pi^2Z(\phi)}\left[\frac{1}{Z^2(\phi)}
\left(\frac{5}{2}\left(Z^{(1)}(\phi)\right)^2-Z(\phi)Z^{(2)}(\phi)\right)
\ln\left(1+\frac{Z(\phi)\Lambda^2}{m^2+U^{(2)}(\phi)}\right)\right.\nonu
&~&~~~~~~~~~~~~~~+\frac{1}{Z^2(\phi)}\left(Z(\phi)Z^{(2)}(\phi)-\frac{43}{12}
\left(Z^{(1)}(\phi)\right)^2\right)\nonu
&~&~~~~~~~~~~~~~~+\left.\frac{1}{Z(\phi)}\frac{Z^{(1)}(\phi)
U^{(3)}(\phi)}{\left(m^2+U^{(2)}(\phi)\right)}
-\frac{1}{6}\frac{\left(U^{(3)}(\phi)\right)^2}
{\left(m^2+U^{(2)}(\phi)\right)^2}\right]
\eea
In the local potential approximation $Z=1$ we obtain
\be
\partial_{m^2}U(\phi)=-{m^2+U^{(2)}(\phi)\over32\pi^2}
\ln\left(1+{\Lambda^2\over m^2+U^{(2)}(\phi)}\right)
\ee
after removing a field independent term. In order to simplify 
the scaling relations we consider the regime $m^2\gg U^{(2)}$, where
\be\label{intas}
\partial_{m^2}U(\phi)=-{1\over32\pi^2}
\ln\left({m^2+\Lambda^2\over m^2}\right)U^{(2)}(\phi)
\ee
Let us come back to the analogy with the infinitesimal
Wilsonian renormalization group method.
In the case of a sharp cutoff $k$, the evolution equation in 
the local potential approximation 
in dimension $d=4$ for $N=1$ is given by (\ref{whlocpot}):
\be
k\partial_kU(\phi)=-{k^4\over16\pi^2}\ln\left(\frac{k^2+U^{(2)}(\phi)}
{k^2+U^{(2)}(\phi_0)}\right)
\ee
which reads in the same regime $k^2\gg U^{(2)}$
\be\label{whas}
k\partial_kU(\phi)=-{k^2\over16\pi^2}U^{(2)}(\phi).
\ee
after removing a field independent term.
The evolutions \eq{intas} and \eq{whas} agree if we impose the differential 
relation between the scale and mass parameters $k$ and $m$
\be
2k\frac{dk}{dm^2}=\ln\left({m^2+\Lambda^2\over m^2}\right)
\label{match}
\ee
We obtain in this manner the usual justification of calling
the Callan-Symanzik equation a renormalization group method
where the mass scale $m$ plays the role of a running cutoff $k$.
The equivalence of the scales and the
elimination of the non-universal contributions requires that the cutoff 
should be far above the mass, $m^2\ll\Lambda^2$.

The non-vanishing anomalous dimension can be recovered with (\ref{diffuz})
In fact, when $Z\not=1$ 
the relation \eq{match} becomes field dependent according to the
first equation of \eq{diffuz}. It is worthwhile comparing what
\eq{diffuz} gives in the asymptotical regime $m^2\gg U^{(2)}$,
\be
\partial_{m^2}Z_{m^2}(\phi)=-\frac{1}{32\pi^2Z_{m^2}^3(\phi)}
\ln\left(\frac{Z_{m^2}(\phi)\Lambda^2+m^2}{m^2}\right)
\left[Z_{m^2}(\phi)Z_{m^2}^{(2)}(\phi)-\frac{5}{2}
\left(Z_{m^2}^{(1)}(\phi)\right)^2\right]
\label{inza}
\ee
with the prediction of the Wegner-Houghton equation.
A possible attempt to save the gradient expansion with sharp cutoff
for the latter is the following: The contributions to the coefficient
functions of the gradient, such as $Z_k(\phi)$, come from
taking the derivative of the loop integral, the trace in (\ref{wheq}),
with respect to the momentum of the
infrared background field $\tilde\phi(x)$. 
There are two kind of contributions,
one which comes form the derivative of the integrand, another
from the external momentum dependence of the limit of the 
integration. It is easy to verify that the $\epsilon$-dependent 
non-local contributions come form the second types only \cite{morr}.
Thus one may consider the approximation where these contributions
are simply neglected, assuming a cancellation mechanism
between the successive blocking steps. The result is, for 
$k^2\gg U^{(2)}_k(\phi)$, c.f. Appendix C,
\be\label{equazk}
k\partial_kZ_k(\phi)=-\frac{k^2}{32\pi^2Z_k^2(\phi)}
\left[2Z_k(\phi)Z_k^{(2)}(\phi)-\frac{5}{2}
\left(Z_k^{(1)}(\phi)\right)^2\right].
\ee
The formal similarity between the two different schemes, \eq{inza} and
\eq{equazk}, can be considered as a measure of the cancellation of 
the non-local terms evoked above.

Finally, we show that we recover the well-known one loop effective action.
For this we consider the solution of (\ref{diffuz}) in the 
independent mode approximation where the $m^2$ dependence 
is ignored in the integrals, $U(\phi)=U_B(\phi)$ and $Z(\phi)=1$.
We get
\bea
U_{eff}(\phi)&=&U_B(\phi)
+\hf\int^0_{M^2}dm^2\int_p
{1\over p^2+m^2+U_B^{(2)}(\phi)}\nonu
&=&U_B(\phi)+\hf\int_p\ln
[p^2+U_B^{(2)}(\phi)]+O(M^{-2}),
\eea
which reproduces the usual one-loop effective potential for
$M\gg\Lambda$. For the kinetic term, the integration of (\ref{diffuz})
in the same approximation leads to
\bea
Z_{eff}(\phi)&=&1-\frac{1}{192\pi^2}\int_{M^2}^0dm^2
\frac{\left(U_B^{(3)}(\phi)\right)^2}{\left(m^2+U_B^{(2)}(\phi)\right)^2}\nonu
&=&1+\frac{1}{192\pi^2}\frac{\left(U_B^{(3)}(\phi)\right)^2}
{U_B^{(2)}(\phi)}+O(M^{-2}),
\eea
for $d=4$ which reproduces the one-loop solution found in \cite{fraser}.
The agreement between the independent mode approximation to our method
and the one-loop solution
is expected because the right hand side of \eq{evolkk} is $\ord(\hbar)$.
This can be understood as a scheme independence of the one loop gamma functions.
But this agreement does not hold beyond $\ord(\hbar)$
as indicated by the imcompatibility of \eq{inza} and \eq{equazk}.

\section{$\hbar$ dependence}
It may happen that the quadratic suppression is not well suited
to a problem. In the case $S_B[\phi]$ possesses local symmetries which should be
preserved then another choice is more appropriate.
The application of our procedure for a gauge model can for
example be based on the choice
\bea
S_B[A]&=&-{1\over4g^2_B}\int dxF_{\mu\nu}^aF^{\mu\nu a}+S_{gf}[A],\nonu
S_s[A]&=&-{1\over4g^2_B}\int dxF_{\mu\nu}^aF^{\mu\nu a},
\eea
where $S_{gf}$ contains the gauge fixing terms and on the application of 
a gauge invariant regularization scheme. As mentioned after eq.
\eq{csrg} we need a regulator to start with in order to follow the dependence
on the amplitude of the fluctuations. One may use lattice, analytic
(asymptotically free models) or Pauli-Villars (QED) regulator to
render \eq{evolk} well defined.
The explicit gauge invariance of $S_s[A]$ which was
achieved by suppressing the gauge covariant field strength instead
of the gauge field itself makes obvious the independence of
the resulting flow for the gauge invariant part of the action
from the choice of the gauge, $S_{gf}$.

We leave the issue of the gauge models for future works and we
return now to the scalar theory and
present the evolution equation for the $\phi^4$ model with quartic
suppression,
\be
S_B[\phi]=S_s[\phi]=\int_x\left[\hf(\partial_\mu\phi_x)^2
+{g_2\over2}\phi_x^2+{g_3\over3!}\phi_x^3+{g_4\over4!}\phi_x^4\right].
\ee
The similarity of this scheme with the loop expansion suggests the
replacement 
\be
{1\over\hbar}=1+\la=1+{1\over g},
\ee
which yields the evolution equation
\be
\partial_g\Gamma[\phi]=-{1\over g^2}e^{-W[j]}
S_s\left[{\delta\over\delta j}\right]e^{W[j]}+{1\over g^2}S_s[\phi].
\ee
The integration of the evolution equation
from $g_{in}=0$ to $g_{fin}=\infty$ corresponds
to the resummation of the loop expansion, i.e. the integration
between $\hbar_{in}=0$ and $\hbar_{fin}=1$.

The gradient expansion ansatz \eq{gradexp} with $Z=1$ gives (c.f. Appendix A)
\bea\label{phin}
\partial_gU(\phi)&=&
-{1\over g^2}\Biggl\{\hf\int_p (p^2+g_2)G(p)\\
&&+{g_3\over3!}\biggl[3\phi\int_pG(p)
-\int_{p_1,p_2}G(p_1)G(p_2)G(-p_1-p_2)
\left(U^{(3)}(\phi)+g^{-1}(g_3+g_4\phi)\right)\biggr]\nonu
&&+{g_4\over4!}\biggl[3\left(\int_pG(p)\right)^2+6\phi^2\int_pG(p)\nonu
&&~~~~~~~-\int_{p_1,p_2,p_3}G(p_1)G(p_2)G(p_3)G(-p_1-p_2-p_3)
\left(U^{(4)}(\phi)+g^{-1}g_4\right)\nonu
&&~~~~~~~-3\int_{p_1,p_2,p_3}G(p_1)G(p_2)G(p_3)G(-p_1-p_2)G(-p_1-p_2-p_3)\nonu
&&~~~~~~~~~~~~~~~~~~~~~~~\times\left(U^{(3)}(\phi)
+g^{-1}(g_3+g_4\phi)\right)^2\nonu
&&~~~~~~~-4\phi\int_{p_1,p_2}G(p_1)G(p_2)G(-p_1-p_2)
\left(U^{(3)}(\phi)+g^{-1}(g_3+g_4\phi)\right)\biggr]\Biggr\},
\nonumber
\eea
where we used the fact that the Fourier transform of the 1PI amplitude 
for $n\ge3$ and $Z=1$ is
\be
\int_{x_1,\cdots,x_n}e^{i(p_1\cdot x_1+\cdots+p_n\cdot x_n)}
\Gamma^{(n)}(x_1,\cdots,x_n)=
(2\pi)^d\delta(p_1+\cdots+p_n)U^{(n)}(\phi).
\ee
The propagator in the presence of the homogeneous background field $\phi$
is given by
\be
G(p)=\left[p^2+U^{(2)}(\phi)+g^{-1}\left(p^2+g_2+g_3\phi
+{g_4\over 2}\phi^2\right)\right]^{-1}.
\ee
Since the momentum dependence in the right hand side of \eq{phin} 
is explicit and simple the one, two and three loop integrals can be carried out
easily by means of the standard methods. The successive derivatives of the
resulting expression with respect to $\phi$ yield the renormalization group
coefficient functions.

The use of our functional equations described in this section 
shows that this method can be generalized to any kind of action $S_s$
and not only to a quadratic suppression term, as shown in the previous
sections.

\section{Summary}
The formal strategy of the renormalization group is generalized in this
paper in such a manner that it includes the Callen-Symanzik scheme and the
resummation of the loop-expansion as two possibilities. These kinds of
generalization depart from the original spirit of the renormalization group
program because the resulting flow does not correspond to the same physics,
instead it interpolates between a suitable chosen, perturbatively
solvable initial condition and the actual bare model. This property is
shared with usual the "exact" renormalization group schemes where
the dependence of the effective action on an IR cutoff is followed.

Our scheme can be considered as a renormalization group method in the
space of the field amplitudes what is usually called the internal space. 
The evolution in the control parameter along the flow
corresponds to the taking into account
the contributions of fluctuations with increasing strength.
Such an iterative inclusion of the fluctuations according to their
scale parameter in the space of the amplitudes, instead of
their scale parameter in the space-time, their momentum, is the
difference between the usual renormalization group procedure
and the ones described in this article.

{\em Note added in proof:} After this work has been completed we learned
that a method presented for gauge models in ref. \cite{simi} is similar
to ours in the case of mass dependence (section IV). \cite{simi} gives
a loop expanded solution of the exact equation, whereas our solution 
is built in the framework of the derivative expansion. Finally,
our approach can be generalized to any kind of suppression action $S_s$
which is compatible with the symmetries as shown in section V.

\begin{appendix}
\section{Legendre transformation}
We collect in this Appendix the relations between 
the derivatives of the generator functional $W[j]$ and
$\Gamma[\phi]$ used in obtaining the evolution equations
for $\Gamma$.

We start with the definitions
\be
W[j]+\Gamma[\phi]+\la S_s[\phi]=j\cdot\phi,
\ee
and
\be
\phi_x=W^{(1)}_x.\label{phj}
\ee
The first derivative of $\Gamma$ gives the
inversion of \eq{phj},
\be
\Gamma^{(1)}_x=j_x-\lambda S_{s,x}^{(1)}.
\label{jph}
\ee
The second derivative is related to the propagator
$W^{(2)}_{x_1,x_2}=G_{x_1,x_2}$ 
\be
\Gamma^{(2)}_{x_1,x_2}={\delta j_{x_1}\over\delta\phi_{x_2}}
-\la S_{s,x_1,x_2}^{(2)}=G_{x_1,x_2}^{-1}-\la S_{s,x_1,x_2}^{(2)}.
\label{propi}
\ee
The third derivative is obtained by differentiating \eq{propi},
\be
\Gamma^{(3)}_{x_1,x_2,x_3}=
-\int_{y_1,y_2,y_3}G^{-1}_{x_1,y_1}G^{-1}_{x_2,y_2}G^{-1}_{x_3,y_3}
W^{(3)}_{y_1,y_2,y_3}-\la S_{s,x_1,x_2,x_3}^{(3)}.
\ee
The inverted form of this equation is
\be
W^{(3)}_{x_1,x_2,x_3}=-\int_{y_1,y_2,y_3}G_{x_1,y_1}G_{x_2,y_2}
G_{x_3,y_3}\left(\Gamma^{(3)}_{y_1,y_2,y_3}+\la S_{s,y_1,y_2,y_3}^{(3)}\right).
\ee
The further derivation gives
\bea
\Gamma^{(4)}_{x_1,x_2,x_3,x_4}&=&
\int_{y_1,y_2,y_3,y_4,z_1,z_2}\biggl[
G^{-1}_{x_1,y_1}G^{-1}_{x_2,y_2}G^{-1}_{x_3,y_3}G^{-1}_{x_4,y_4}
W^{(4)}_{y_1,y_2,y_3,y_4}\nonu
&&+G^{-1}_{x_1,y_1}G^{-1}_{x_2,y_2}W^{(3)}_{y_1,y_2,z_1}
G^{-1}_{z_1,z_2}W^{(3)}_{z_2,y_3,y_4}
G^{-1}_{x_3,y_3}G^{-1}_{x_4,y_4}\nonu
&&+G^{-1}_{x_3,y_3}G^{-1}_{x_2,y_2}W^{(3)}_{y_3,y_2,z_1}
G^{-1}_{z_1,z_2}W^{(3)}_{z_2,y_1,y_4}
G^{-1}_{x_1,y_1}G^{-1}_{x_4,y_4}\nonu
&&+G^{-1}_{x_1,y_1}G^{-1}_{x_4,y_4}W^{(3)}_{y_1,y_4,z_1}
G^{-1}_{z_1,z_2}W^{(3)}_{z_2,y_3,y_2}
G^{-1}_{x_3,y_3}G^{-1}_{x_2,y_2}\biggr]\nonu
&&-\la S_{s,x_1,x_2,x_3,x_4}^{(4)}.
\eea
Its inversion expresses the four point connected Green
function in terms of the 1PI amplitudes,
\bea
W^{(4)}_{x_1,x_2,x_3,x_4}
&=&\int_{y_1,y_2,y_3,y_4,z_1,z_2}\biggl[
G_{x_1,y_1}G_{x_2,y_2}G_{x_3,y_3}G_{x_4,y_4}
\left(\Gamma^{(4)}_{y_1,y_2,y_3,y_4}+\la S_{s,y_1,y_2,y_3,y_4}^{(4)}\right)\\
&&-G_{x_1,y_1}G_{x_2,y_2}\left(\Gamma^{(3)}_{y_1,y_2,z_1}+\la 
S_{s,y_1,y_2,z_1}^{(3)}\right)
G_{z_1,z_2}\left(\Gamma^{(3)}_{z_2,y_3,y_4}+\la S_{s,z_2,y_3,y_4}^{(3)}\right)
G_{x_3,y_3}G_{x_4,y_4}\nonu
&&-G_{x_3,y_3}G_{x_2,y_2}\left(\Gamma^{(3)}_{y_3,y_2,z_1}+\la 
S_{s,y_3,y_2,z_1}^{(3)}\right)
G_{z_1,z_2}\left(\Gamma^{(3)}_{z_2,y_1,y_4}+\la S_{s,z_2,y_1,y_4}^{(3)}\right)
G_{x_1,y_1}G_{x_4,y_4}\nonu
&&-G_{x_1,y_1}G_{x_4,y_4}\left(\Gamma^{(3)}_{y_1,y_4,z_1}+\la 
S_{s,y_1,y_4,z_1}^{(3)}\right)
G_{z_1,z_2}\left(\Gamma^{(3)}_{z_2,y_3,y_2}+\la S_{s,z_2,y_3,y_2}^{(3)}\right)
G_{x_3,y_3}G_{x_2,y_2}\biggr].\nonumber
\nonumber
\eea

\section{Evolution equation derivation}
We give here some details on the computation of (\ref{evoluz}).
To get the evolution equation of the potential part of the gradient expansion
(\ref{gradexp}), one has to take a homogeneous field $\phi=\phi_0$
in (\ref{evolkk}). But to distinguish the kinetic contribution from the 
potential one, a non homogeneous field $\phi(x)=\phi_0+\eta(x)$ is needed, as
well. Let $k$ be the momentum where the field $\eta$ is non-vanishing. 
Then the effective action can be written as
\be
\Gamma[\phi]=V_dU_\la(\phi_0)+\hf\int_q\tilde\eta(q)\tilde\eta(-q)
\left(Z_\la(\phi_0)q^2+U_\la^{(2)}(\phi_0)\right)+\ord(\tilde\eta^3,k^4)
\ee
where $V_d$ is the spatial volume. Thus we need the second derivative of 
the effective action in (\ref{evolkk}) up to the second order in 
$\tilde\eta$ to identify the different contributions. The terms 
independent of $\tilde\eta$
give the equation for $U_\la$ and the ones proportional to $k^2\tilde\eta^2$
the equation for $Z_\la$. The contributions proportional to $\tilde\eta^2$ but 
independent of $k$ yield an equation for $U_\la^{(2)}$ which must
be consistent with the equation for $U_\la$. The result is 
\bea
\Gamma^{(2)}_{p_1,p_2}&=&\left[Z_\la(\phi_0)p_1^2+U_\la^{(2)}(\phi_0)\right]
\delta(p_1+p_2)\\
&+&\int_q\tilde\eta(q)\left[Z_\la^{(1)}(\phi_0)(p_1^2+q^2+qp_1)+U_\la^{(3)}(\phi
_
0)\right]
\delta(p_1+p_2+q)\nonu
&+&\hf\int_{q_1,q_2}\tilde\eta(q_1)\tilde\eta(q_2)
\left[Z_\la^{(2)}(\phi_0)(p_1^2+2q_1^2+q_1q_2+2q_1p_1)+U_\la^{(4)}(\phi_0)\right
]
\delta(p_1+p_2+q_1+q_2)\nonu
&+&O(\tilde\eta^3,k^4)\nonumber
\eea
Finally one computes the inverse of the operator 
$\la\cm_{p_1,p_2}+\Gamma^{(2)}_{p_1,p_2}$ and expands it in powers of 
$\tilde\eta$ and $k$. The trace over $p_1$ and $p_2$ needs the 
computations of terms like
\be
\mbox{Tr}\left\{(p_1q_1)(p_2q_2)F(p_1,p_2)\delta(p_1+p_2+q_1+q_2)\right\}=
\frac{q_1^2}{d}\delta(q_1+q_2)\int_p p^2F(p,-p)
\ee
and they lead to (\ref{evoluz}). The consistency with the 
equation for $U_\la^{(2)}$ is satisfied.

\section{Exact renormalization group equations}
For the derivation of the flow equations for the potential $U_k$ and the wave
function renormalization $Z_k$ we will take in the Wegner-Houghton equation 
(\ref{wheq})
a non homogeneous background field $\phi=\phi_0+\phi_l$ where
$\phi_0$ is homogeneous and $\phi_l=\sum_{|q|=l}\phi_q e^{iqx}$.
The expansion of the running action $\Gamma_k$ (in the sens of the 
renormalization
group transformations) in powers of $\phi_l$ gives

\be\label{mg}
\partial_k S_k[\phi_0+\phi_l]=V_d\left(\partial_k U_k(\phi_0)+
\frac{\Phi^2}{2}\left[l^2\partial_k Z_k(\phi_0)+
\partial_k U_k^{(2)}(\phi_0)\right]+...
\right)
\ee
where $\Phi^2=\sum_q\phi_q\phi_{-q}$, so that the expansion of (\ref{wheq})
in powers of $\phi_l$ will help us identify:

\begin{itemize}
\item the evolution of $U_k$ given by the terms independent of $\Phi^2$,

\item the evolution of $Z_k$ given by the terms proportional to $l^2\Phi^2$,

\item the evolution of $U^{(2)}_k$ given by the terms proportional to $\Phi$ 
\end{itemize}
but independent of 
$l$. This equation will of course have to be consistent with the one for $U_k$. 
For this consistency condition, we will need to take $l<<k$, as will be shown.

The Wegner-Houghton equation needs the second derivative of the action which 
reads in the second order in $\phi_l$ as
\bea\label{dss}
\frac{1}{V_d}\frac{\partial^2 S_k}
{\partial\tilde\phi_{p_1}\partial\tilde\phi_{-p_2}}|_{\phi_0+\phi_B}&=&
\left[p_1^2Z_0+U^{(2)}_0\right]\delta(p_1-p_2)\\
&+&\sum_q\tilde\phi_q\left[\left(p_1^2+q^2+qp_1\right)Z_0^{(1)}+U_0^{(3)}
\right]\delta
(p_1-p_2+q)\nonumber\\
&+&\frac{1}{2}\sum_{q_1,q_2}\tilde\phi_{q_1}\tilde\phi_{q_2}\left[
\left(p_1^2+2q_1^2+q_1
q_2+2q_1p_1\right)Z_0^{(2)}
+U_0^{(4)}\right]\delta(p_1-p_2+q_1+q_2).\nonumber
\eea
The expansion of the logarithm in the same order reads

\bea\label{logds}
&~&\ln\left[\frac{1}{V_d}\frac{\partial^2S_k}
{\partial\tilde\phi_{p_1}\partial\tilde\phi_{p_2}}
\right]~=~
\delta(p_1-p_2)\ln\left[p_1^2Z_0+U^{(2)}_0\right]\\
&+&\frac{1}{p_1^2Z_0+U^{(2)}_0}
\sum_q\delta(p_1-p_2+q)\tilde\phi_q\left[\left(p_1^2+q^2+qp_1\right)Z_0^{(
1)}+U_0^{(3)}\right]
\nonumber\\
&+&\frac{1}{2}\frac{1}{p_1^2Z_0+U^{(2)}_0}
\sum_{q_1,q_2}\delta(p_1-p_2+q_1+q_2)\tilde\phi_{q_1}\tilde\phi_{q_2}\Bigg[
\left[\left(p_1^2+2q_1^2+q_1q_2+2q_1p_1\right)Z_0^{(2)}+U_0^{(4)}\right]
\nonumber\\
&-&\frac{1}{(p_2-q_2)^2Z_0+U^{(2)}_0}
\left[\left(p_1^2+q_1^2+q_1p_1\right)Z_0^{(1)}+U_0^{(3)}\right]
\left[\left(p_2^2+q_2^2-q_2p_2\right)Z_0^{(1)}+U_0^{(3)}\right]
{\cal C}(p_1-q_1)\Bigg]\nonumber
\eea
where ${\cal C}(p_1-q_1)$ represents the constraint that $|p_1-q_1|$, as well as
$|p_1|$, must be between $k$ and $k-\delta k$. This is the origin of the constraint 
$l<<k$: if this is not satisfied,
the term proportional to ${\cal C}(p_1-q_1)$ in (\ref{logds}) does not contribute to 
the evolution equations (it is of the order $\delta k^2$) and the evolution of 
$U^{(2)}_k$ is not consistent with the one of $U_k$.

Then we need to expand (\ref{logds})
in powers of $l$. But there appears terms proportional to the first power of $l$ 
which would give
non local contributions to the gradient expansion since they would be written 
$\sqrt\Box$.
There are two ways to rid of the non-local contributions when the model is
solved by the loop expansion, i.e. by means of loop integrals for momenta
$0\le p\le\Lambda$. One is to use lattice regularization where the
periodicity in the Brillouin zone cancel the $q$ dependence of
the domain of the integration. Another way to eliminate the non-local
terms is to remove the cutoff. Since the non-local contributions
represent surface terms they vanish as $\Lambda\to\infty$.

One may furthermore speculate that some of the non-local terms cancel
between the consecutive steps of the blocking $k\to k-\dk$
for a suitable choice of the cutoff function $f(\kappa)$
in the propagator $G^{-1}_k(p)=f(p/k)G^{-1}(p)$.
Ignoring simply the non-local terms the identification of the
coefficients of the different powers in the derivative expansion, we 
finally obtain from the 
Wegner-Houghton equation

\bea
k\partial_kU_k(\phi_0)&=&-\frac{\hbar\Omega_d k^d}{2(2\pi)^d}
\ln\left(\frac{Z_k(\phi_0)k^2+U_k^{(2)}(\phi_0)}
{Z_k(0)k^2+U_k^{(2)}(0)}\right)\\
k\partial_kZ_k(\phi_0)&=&-\frac{\hbar\Omega_d k^d}{2(2\pi)^d}\left(
\frac{Z_k^{(2)}(\phi_0)}{Z_k(\phi_0)k^2+U_k^{(2)}(\phi_0)}-2Z_k^{(1)}(\phi_0)
\frac{Z_k^{(1)}(\phi_0)k^2+U_k^{(3)}(\phi_0)}
{\left(Z_k(\phi_0)k^2+U_k^{(2)}(\phi_0)\right)^2}\right.\nonumber\\
&-&\frac{k^2}{d}\frac{\left(Z_k^{(1)}(\phi_0)\right)^2}
{\left(Z_k(\phi_0)k^2+U_k^{(2)}(\phi_0)\right)^2}
+\frac{4k^2}{d}Z_k(\phi_0)Z_k^{(1)}(\phi_0)
\frac{Z_k^{(1)}(\phi_0)k^2+U_k^{(3)}(\phi_0)}
{\left(Z_k(\phi_0)k^2+U_k^{(2)}(\phi_0)\right)^3}\nonumber\\
&+&\left.Z_k(\phi_0)
\frac{\left(Z_k^{(1)}(\phi_0)k^2+U_k^{(3)}(\phi_0)\right)^2}
{\left(Z_k(\phi_0)k^2+U_k^{(2)}(\phi_0)\right)^3}
-\frac{4k^2}{d}Z_k^2(\phi_0)
\frac{\left(Z_k^{(1)}(\phi_0)k^2+U_k^{(3)}(\phi_0)\right)^2}
{\left(Z_k(\phi_0)k^2+U_k^{(2)}(\phi_0)\right)^4}\right)
\nonumber
\eea
where the origin of the potential has been chosen at $\phi_0=0$.

When $k^2\gg U^{(2)}_k(\phi)$ this gives \eq{equazk} in dimension $d=4$.

\end{appendix}


\end{document}